\documentclass[aps,prl,twocolumn,showpacs,floatfix]{revtex4}

\usepackage{dcolumn}
\usepackage{amssymb}
\usepackage{graphicx}

\begin{document}

\newcommand \be {\begin{equation}}
\newcommand \ee {\end{equation}}
\newcommand \bea {\begin{eqnarray}}
\newcommand \eea {\end{eqnarray}}
\newcommand \nn {\nonumber}
\newcommand \la {\langle}
\newcommand \rl {\rangle_L}
\newcommand \ve {\varepsilon}

\title{Far-from-equilibrium state in a weakly dissipative model}
\author{Eric Bertin$^1$ and Olivier Dauchot$^2$}
\affiliation{$^1$ Universit\'e de Lyon, Laboratoire de Physique,
Ecole Normale Sup\'erieure de Lyon, CNRS,
46 all\'ee d'Italie, F-69007 Lyon, France\\
$^2$ Service de Physique de l'Etat Condens\'e, CEA Saclay,
F-91191 Gif-sur-Yvette Cedex}

\date{\today}

\begin{abstract}
We address, on the example of a simple solvable model,
the issue of whether the stationary state of dissipative systems
converges to an equilibrium state in the low dissipation limit.
We study a driven dissipative
Zero Range Process on a tree, in which particles are
interpreted as finite amounts of energy exchanged between degrees
of freedom. The tree structure mimicks the hierarchy of length scales;
energy is injected at the top of the tree ('large scales'),
transferred through the tree and dissipated mostly in the deepest branches
of the tree ('small scales').
Varying a parameter characterizing the transfer dynamics, a transition
is observed, in the low dissipation limit, between a quasi-equilibrated
regime and a far-from-equilibrium one, where the dissipated flux
does not vanish.
\end{abstract}


\maketitle

One of the main challenges of non-equilibrium statistical physics is to
understand which principles rule the description of nonequilibrium
stationary states. Generic approaches, like linear response theory
\cite{Kubo95}, have been developed for weakly driven systems
(for instance gently stirred fluids, or systems subjected to small
temperature gradients).
In this situation, when the external forcing is very small,
the system remains close to an equilibrium state, and
the effect of the small drive is thus perturbative.
A different situation, which also attracted a lot of interest,
is that of dissipative systems, in which an energy flux takes place
between scales rather than in real space. Energy is usually injected
at large scales, and cascades down, through non-linear interactions,
to smaller scales where it is dissipated.
Examples include hydrodynamic turbulence \cite{Kraichnan89,Frisch95},
wave turbulence in fluids or plasma \cite{Zakharov92},
ferrofluids \cite{Boyer08} and vibrating plates \cite{During06,Boudaoud08},
fracture \cite{fracture} and friction \cite{Volmer97},
as well as granular materials \cite{Haff83,Jaeger96} and foams
\cite{Lundberg08}.
Dissipative effects are usually characterized by a dissipation parameter,
like viscosity or inelasticity coefficients, which is zero
in the conservative case.
A natural question is to know whether in the limit of small,
but nonzero dissipation coefficients,
the stationary state of the system becomes
close to some equilibrium state to be determined.
Qualitatively, one may expect that adding a tiny amount of injection
and dissipation to a conservative system breaks energy conservation,
but leads to small fluctuations around a given energy level selected
by the injection and dissipation mechanisms. The system would thus
merely behave as if it was at equilibrium at this energy,
and a perturbative approach around an equilibrium state may be meaningful.

Whether this scenario holds in general is certainly an open issue.
Perturbative approaches around equilibrium states for dissipative
systems have been proposed in the context of two-dimensional turbulence
\cite{Montgomery74,Miller90,Robert91,Chavanis00,Dubrulle06,Bouchet08},
for which the flux
of dissipated energy vanishes in the small viscosity limit \cite{Frisch95}.
However, in other situations like three-dimensional turbulence
\cite{Frisch95} or granular gases \cite{Aumaitre01}, the dissipated flux
seems to remain finite for small viscosity, suggesting that the statistical
state of the system does not converge to any equilibrium state.

\begin{figure}[b]
\centering\includegraphics[height=4cm,clip]{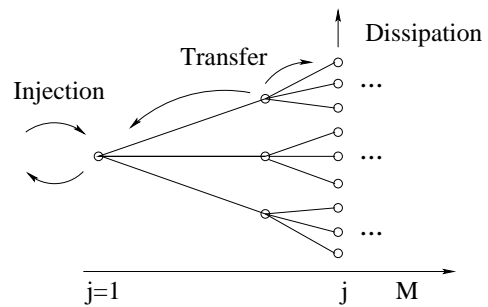}
\caption{\sl Sketch of the model, illustrating the tree geometry.}
\label{fig-tree}
\end{figure}

In order to give further insights into these issues,
simple solvable models may be helpful. Here, we study 
a stochastic transport model, namely a Zero Range Process
\cite{Hanney05,Noh05}, that
describes in a schematic way the transfer of energy between scales,
in the presence of injection and dissipation.
To account for the specific organization of scale space, where small
scales are much more ``numerous'' than large scales, we define our
model on a tree geometry. 
We show that depending on the energy transfer dynamics, the dissipative
stationary state of the model converges in the weak dissipation limit
either to an equilibrium state, or to a well-defined far-from-equilibrium
state with a finite dissipated flux.

The model is defined on a tree composed of $M$ successive levels
(see Fig.~\ref{fig-tree});
at any given level $j<M$, all sites have $m>1$ forward branches that link
them to level $j+1$, so that the number of sites at level $j$ is $m^{j-1}$.
Sites are thus labeled by the level index $j$, and the index
$i=1,\ldots,m^{j-1}$ within level $j$.
The energy on each site $(j,i)$ of the tree is assumed to take only
discrete values proportional to an elementary amount $\ve_0$,
namely $\ve_{j,i}=n_{j,i} \ve_0$.
Energy transfer proceeds as follows: an energy amount $\ve_0$
is moved, either forward or backward, along any branch between
levels $j$ and $j+1$ with a rate $\nu_j$.
In the absence of driving and dissipation,
the dynamics satisfies detailed balance.
Energy injection is implemented by connecting the site $(1,1)$
to a thermostat of temperature $T_\mathrm{ext}=\beta_\mathrm{ext}^{-1}$,
with a coupling frequency $\nu_\mathrm{ext}$.
Dissipation proceeds through the random withdrawal of an amount of energy
$\ve_0$ at site $(j,i)$ with rate $\Delta_j$.
The master equation governing the evolution of the probability
distribution $P(\{n_{j,i}\},t)$ reads:

\begin{widetext}
\bea \nonumber
\frac{\partial P}{\partial t}(\{n_{j,i}\},t) &=& -\left( \sum_{j=1}^{M-1}
m^{j-1}(2m\nu_j+\Delta_j) + m^{M-1} \Delta_M + \nu_\mathrm{ext} \left(
1+e^{-\beta_\mathrm{ext}\ve_0}\right) \right) P(\{n_{j,i}\},t)\\ \nonumber
&+& \sum_{j=1}^{M-1} \sum_{i=1}^{m^{j-1}} \sum_{l=1}^m \nu_j
\left[ P(n_{j,i}+1,n_{j+1,(i-1)m+l}-1,\{n_{q,r}\},t)
+ P(n_{j,i}-1,n_{j+1,(i-1)m+l}+1,\{n_{q,r}\},t)\right]\\
&+& \sum_{j=1}^M \sum_{i=1}^{m^{j-1}} \Delta_j P(n_{j,i}+1,\{n_{q,r}\},t)
+ \nu_\mathrm{ext} \left[ P(n_{1,1}+1,\{n_{q,r}\},t)
+ e^{-\beta_\mathrm{ext}\ve_0} P(n_{1,1}-1,\{n_{q,r}\},t)\right].
\label{ME}
\eea
\end{widetext}
In this last equation, $\{n_{q,r}\}$ is a short-hand notation
for all the variables that are not explicitely listed.
Turning to the stationary state,
we look for a steady-state probability distribution of the form:
\be \label{st-form}
P_\mathrm{st}(\{n_{j,i}\}) = \frac{1}{Z} \prod_{j=1}^M \prod_{i=1}^{m^{j-1}}
e^{-\beta_j n_{j,i} \ve_0}
\ee
where $\beta_j$ is an 'effective' inverse temperature (to be determined)
associated to level $j$, and $Z$ is a normalization factor.
Inserting expression (\ref{st-form}) of the stationary distribution
into the master equation (\ref{ME}) yields a set of equations
to be satisfied by the parameters $z_j=\exp(-\beta_j \ve_0)$,
for $j=2,\ldots,M-1$:
\be
\label{eq-zj}
\nu_{j-1} (z_{j-1}-z_j) - m\nu_j (z_j-z_{j+1}) = \Delta_j z_j
\ee
with the boundary conditions
\bea
\label{eq-z1}
\nu_\mathrm{ext} (e^{-\beta_\mathrm{ext}\ve_0}- z_1)
-m\nu_1 (z_1-z_2) &=& \Delta_1 z_1\\
\label{eq-zM}
\nu_{M-1} (z_{M-1} - z_M) &=& \Delta_Mz_M.
\eea
Note that these equations correspond to the local balance of the diffusive
fluxes $\nu_j (z_j-z_{j+1})$ and dissipative fluxes $\Delta_j z_j$.
In the absence of dissipation, namely if $\Delta_j=0$ for all $j$,
the equilibrium solution $\beta_1=\ldots=\beta_M=\beta_\mathrm{ext}$
is recovered.
To study the dissipative case, we need to choose a specific form of 
the frequency $\nu_j$ and the dissipation rate $\Delta_j$.
A generic form is the following:
\be \label{Delta-gamma}
\nu_j = \nu_0 k_j^{\alpha}, \qquad
\Delta_j = D k_j^{\gamma}, \qquad \gamma >0,
\ee
where we have introduced a pseudo-wavenumber $k_j=m^{j-1}$,
to map the tree structure onto physical space.
Parameters $\nu_0$ and $D$ are respectively a frequency characterizing
the large scale dynamics, and a dissipation coefficient.
We impose the condition $\alpha<\gamma$, so that
dissipation becomes the dominant effect at small scales (large $k_j$).
The transfer rate $\nu_j$ and the dissipation rate $\Delta_j$ are
balanced for a wavenumber $k_j=K$ given by
\be \label{def-K}
K = \left(\frac{\nu_0}{D}\right)^{1/(\gamma-\alpha)},
\ee
which goes to infinity in the limit of small dissipation coefficient $D$.
Note that $\nu_0/D$ is similar to the Reynolds number in hydrodynamics.
For large $K$, we shall call the ranges $k_j \ll K$ and $k_j \gg K$
the ``inertial'' and ``dissipative'' ranges respectively.

\begin{figure}[t]
\centering\includegraphics[width=8cm,clip]{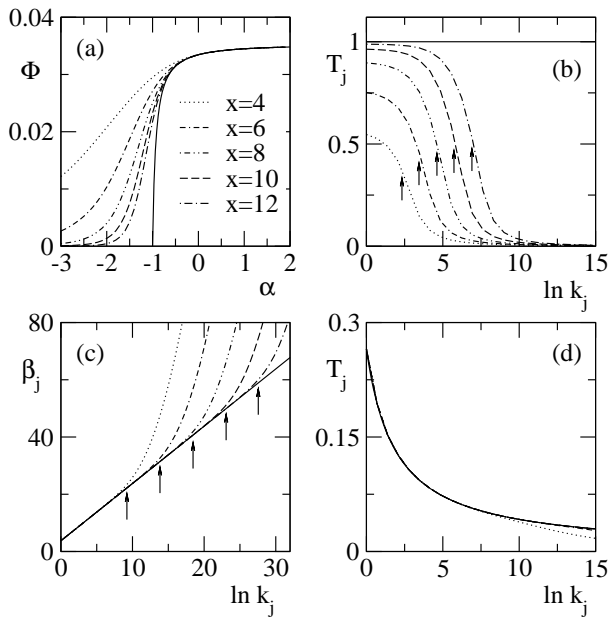}
\caption{\sl Numerical solution of
Eqs.~(\ref{eq-zj}), (\ref{eq-z1}) and (\ref{eq-zM})
for $\gamma=2$ and $D=10^{-x}$, with $x$
given in (a); same symbols for all four figures.
(a) Energy flux $\Phi$ as a function of $\alpha$;
the full line is $\Phi_0$ given in Eq.~(\ref{eq-Phi0}).
(b) Temperature $T_j=\beta_j^{-1}$ versus $\ln k_j$ for $\alpha=-2$;
the full line is $T_\mathrm{ext}$.
(c) $\beta_j$ versus $\ln k_j$ for $\alpha=1$;
full line: $\beta_j^\mathrm{neq}$ defined in Eq.~(\ref{eq-betaj-he}).
(d) Same data as (c) plotted as $T_j$, on the same
window of $\ln k_j$ as (b).
Arrows in (b) and (c) indicate the value of $K$ for each $D$.
Other parameters: $M=50$, $m=2$, $\nu_0=1$,
$\nu_\mathrm{ext}=0.1$, $\beta_\mathrm{ext}=1$, $\ve_0=1$.}
\label{fig-Tj}
\end{figure}

The solution of Eqs.~(\ref{eq-zj}), (\ref{eq-z1}) and (\ref{eq-zM})
can be evaluated numerically.
We are interested in the inertial range bahavior,
where energy transfer dominates over
dissipative effects, so that we shall explore the solutions by varying
$\alpha$ while keeping $\gamma$ fixed.
We first compute the mean energy flux $\Phi$ injected by the reservoir,
\be \label{def-Phi}
\Phi = \nu_\mathrm{ext} \left( e^{-\beta_\mathrm{ext}\ve_0}
-e^{-\beta_1\ve_0}\right).
\ee
The flux $\Phi$ is plotted as a function of $\alpha$ in Fig.~\ref{fig-Tj}(a),
for a broad range of values of the dissipation coefficient $D$.
We observe a transition around the value $\alpha=-1$: for $\alpha<-1$,
$\Phi \to 0$ when $D\to 0$, while for $\alpha >-1$, $\Phi$
converges to a finite value in the small $D$ limit.
These two regimes are also clearly seen in Fig.~\ref{fig-Tj} (b), (c), and (d)
by plotting the temperature $T_j = \beta_j^{-1} = -\ve_0/\ln z_j$
as a function of $\ln k_j=(j-1)\ln m$.
A first trivial observation is that $K$ increases more rapidly when
decreasing $D$ for larger values of $\alpha$.
More interestingly, we observe that for $\alpha=-2$ [Fig.~\ref{fig-Tj}(b)],
the temperature profile slowly converges
to the equilibrium profile $T_j^{eq}=\beta_\mathrm{ext}^{-1}$ when $D \to 0$,
while for $\alpha=1$ [Fig.~\ref{fig-Tj}(c) and (d)], it
converges to a well-defined nonequilibrium profile,
which is linear for $k_j \lesssim K$
when plotting $\beta_j$ as a function of $\ln k_j$
[see Fig.~\ref{fig-Tj}(c)].
These results can be interpreted as follows. 
When the transfer mechanism is inefficient at small scales ($\alpha <-1$),
dissipative scales are not ``feeded'', so that energy accumulates at large
scales, generating an effective equilibrium.
In the opposite case ($\alpha >-1$), the transfer mechanism
is efficient at small scales, thus ``pumping'' energy from large scales
to dissipative ones.

Most of the above behavior can be understood using a simpler form
of the dissipation, which leads to analytically tractable calculations.
We assume that $\Delta_j=0$ for all $j<M$, leaving a nonzero dissipation
rate $\Delta_M$ only on the last level $j=M$ of the tree.
As a first step, we look for the solutions of
Eqs.~(\ref{eq-zj}) and (\ref{eq-z1}) with $\Delta_j=0$, $j=1,\ldots,M-1$,
without taking into account the dissipative boundary condition
(\ref{eq-zM}). We find a family of solutions, parameterized by the flux $\Phi$:
\be \label{eq-zj2}
z_j = e^{-\beta_\mathrm{ext}\ve_0} \left[ 1- \Phi
\left( \frac{1}{\Phi_0}-\frac{1}{B\, k_j^{1+\alpha}} \right) \right],
\ee
where $j=1,\ldots,M$, and with $\Phi_0$ and $B$ given by
\bea \label{eq-Phi0}
\Phi_0 &=& \frac{\nu_0 (m-m^{-\alpha})\nu_\mathrm{ext}\, e^{-\beta_\mathrm{ext}\ve_0}}{\nu_0 (m-m^{-\alpha})+\nu_\mathrm{ext}},\\
B &=& \nu_0 (m-m^{-\alpha})\, e^{-\beta_\mathrm{ext}\ve_0}.
\eea
Note that $z_j$ smoothly converges to $e^{-\beta_\mathrm{ext}\ve_0}$
when $\Phi \to 0$, and is a decreasing function of $k_j$
for $\Phi>0$.
Interestingly, Eq.~(\ref{eq-zj2}) imposes an upper bound $\Phi_\mathrm{max}$
on the flux $\Phi$, which is determined by the condition $z_M>0$:
\be
\Phi_\mathrm{max}=\left(\frac{1}{\Phi_0}-
\frac{1}{B\,k_M^{1+\alpha}}\right)^{-1}.
\ee
If $\alpha<-1$, one finds for large $M$ that
$\Phi_\mathrm{max} \approx |B|\, k_M^{-|1+\alpha|}$, so that
$\Phi_\mathrm{max} \to 0$ when $M\to \infty$.
Accordingly, whatever the small scale boundary condition,
the flux vanishes in the large size limit.
In contrast, if $\alpha>-1$, $\Phi_\mathrm{max}$ converges to $\Phi_0>0$
in the large size limit $M\to \infty$; $\Phi_0$ goes to zero linearly
with $\alpha$ when $\alpha \to -1^{+}$.

We now use the dissipative boundary condition (\ref{eq-zM})
to determine the precise value of the flux.
Eq.~(\ref{eq-zM}) states that the diffusive flux $\Phi$ is equal to
the dissipated flux on level $j=M$.
This condition leads to $\Phi = m^{M-1} \Delta_M z_M$,
or using $k_M=m^{M-1}$, $z_M=\Phi/(k_M \Delta_M)$.
Equating this value of $z_M$ with that given in Eq.~(\ref{eq-zj2})
for $j=M$, yields an equation for $\Phi$, which is solved into
\be \label{eq-Phi}
\Phi = \Phi_\mathrm{max}
\left( 1+\frac{e^{\beta_\mathrm{ext}\ve_0}\Phi_\mathrm{max}}{k_M \Delta_M}
\right)^{-1}.
\ee
Identifying $k_M$ with the value $K$ defined in Eq.~(\ref{def-K}),
we get $\Delta_M=D K^{\gamma}=\nu_0 K^{\alpha}$, and thus
$k_M \Delta_M =\nu_0 K^{1+\alpha}$.
For $\alpha<-1$, both $k_M \Delta_M$ and $\Phi_\mathrm{max}$ are
for large $K$ proportional to $K^{-|1+\alpha|}$, so that their ratio
is a constant. From Eq.~(\ref{eq-Phi}), $\Phi$ goes to zero as
a finite fraction of the maximum flux $\Phi_\mathrm{max}$.
Using $K\sim D^{-1/(\gamma-\alpha)}$, the flux $\Phi$ behaves
in terms of the dissipation coefficient as $\Phi \sim D^{\mu}$
when $D \to 0$, with $\mu = |1+\alpha|/(\gamma-\alpha)$.
From Eq.~(\ref{eq-zj2}), $\beta_j \to \beta_\mathrm{ext}$ when
$D \to 0$, as long as $k_j \ll K$.
Altogether, the effect of dissipation on the system may be considered
as perturbative in the case $\alpha<-1$.
The perturbation expansion is however singular, with a nontrivial
exponent $\mu<1$.
In the opposite case $\alpha>-1$, $\Phi_\mathrm{max} \to \Phi_0>0$,
while $k_M \Delta_M \to \infty$. Hence from Eq.~(\ref{eq-Phi}),
$\Phi$ is equal for large $K$ (or small $D$) to the maximum flux
$\Phi_\mathrm{max} = \Phi_0$, consistently with Fig.~\ref{fig-Tj}(a)
\footnote{Similar results are obtained if one chooses a constant value
$\Delta_M = \Delta_0$ instead of $\Delta_M = \nu_0 K^{\alpha}$.}.
From Eq.~(\ref{eq-zj2}),
the temperature profile $\beta_j=-\ve_0^{-1}\ln z_j$ converges
to a well-defined nonequilibrium profile
\be \label{eq-betaj-he}
\beta_j^{\mathrm{neq}} = \frac{1}{\ve_0}(1+\alpha)\ln k_j + \beta_\mathrm{ext}
+\frac{1}{\ve_0}\ln C,
\ee
with $C=1+\nu_0 (m-m^{-\alpha})/\nu_\mathrm{ext}$.
Although this profile has been obtained with a simplified version of the
model, one sees on Fig.~\ref{fig-Tj}(c) that $\beta_j$ computed numerically
in the original model also converges to $\beta_j^{\mathrm{neq}}$.
Let us emphasize that $\beta_j^{\mathrm{neq}}$ does not depend on
parameters related to dissipation \footnote{Obviously, $\beta_j^{\mathrm{neq}}$
cannot depend on $D$ as the limit $D\to 0$ is taken, but it could depend
on $\gamma$.}, but only on parameters characterizing injection and transfer.
The temperature profile is continuous with $\alpha$, namely
$\beta_j \to \beta_\mathrm{ext}$ when $\alpha \to -1^{+}$.
Note also that the coupling $\nu_\mathrm{ext}$
simply ``renormalizes'' the inverse temperature $\beta_\mathrm{ext}$;
the low coupling limit corresponds to driving the system with
a small effective temperature.

\begin{figure}[t]
\centering\includegraphics[width=8cm,clip]{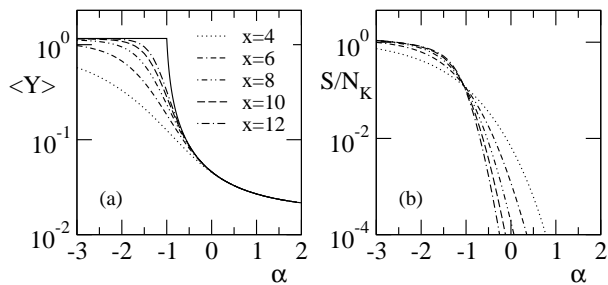}
\caption{\sl (a) $\langle Y\rangle$ as a function of $\alpha$,
for $D=10^{-x}$; $Y$ is defined in Eq.~(\ref{eq-Y}), with here
$f(n)=n$ and $g(k)=k^{-2}$.
The full line is the asymptotic value of $\langle Y\rangle$ for $D\to 0$.
(b) Normalized entropy $S/N_K$ versus $\alpha$, for the same values of $D$
as (a). Other parameters: same as in Fig.~\ref{fig-Tj}.
}
\label{fig-observ}
\end{figure}

To sum up, it turns out that an equilibrium approach to the 
stationary state of the present model in the weak dissipation limit
is meaningful if $\alpha <-1$.
In this case, the probability distribution $P_\mathrm{st}(\{n_{j,i}\})$
converges, in a weak sense, to the equilibrium distribution of temperature
$\beta_\mathrm{ext}$.
Although some deviations from equilibrium persist at small scale
in the distribution
for $D \to 0$, the average value of ``large scale'' observables
converges to the corresponding equilibrium value.
Such ``large scale'' quantities, which are not
sensitive to small scale details of the distribution,
include observables $Y$ defined as
\be \label{eq-Y}
Y = \sum_{j=1}^M \left( g(k_j) \sum_{i=1}^{m^{j-1}} f(n_{j,i}) \right),
\ee
where $f(n)$ is an arbitrary function, and $g(k)$ satisfies $kg(k) \to 0$
when $k\to \infty$.
In the opposite case $\alpha>-1$, the probability distribution
$P_\mathrm{st}(\{n_{j,i}\})$ converges when $D\to 0$
to a well-defined nonequilibrium
probability distribution $P_\mathrm{neq}(\{n_{j,i}\})$, given by
Eqs.~(\ref{st-form}) and (\ref{eq-betaj-he})
\footnote{The zero dissipation limit should however be taken after
the infinite size limit; otherwise, equilibrium is recovered.}.
The convergence of $\langle Y \rangle$ to
$\langle Y \rangle_\mathrm{eq}$ for $\alpha<-1$ and to
$\langle Y \rangle_\mathrm{neq}$ for $\alpha>-1$ is illustrated
on an example in Fig.~\ref{fig-observ}(a).
Note that $\langle Y \rangle_\mathrm{neq}$ depends on $\alpha$,
and thus on the energy transfer, while
$\langle Y \rangle_\mathrm{eq}$ obviously does not.

Other relevant statistical quantities are sensitive to the small scale
details of the distribution, and have a more complex behavior.
This is the case of the entropy
\be
S=-\sum_{\{n_{j,i}\}} P_{st}(\{n_{j,i}\}) \ln P_{st}(\{n_{j,i}\}).
\ee
As for $\alpha<-1$, $T_j$ drops from $T_\mathrm{ext}$ to zero
for $k_j \approx K$,
it is natural to expect that the entropy $S$ is proportional at large
$K$ to the number $N_K$ of sites $(j,i)$ with $k_j \le K$, namely
$N_K = mK/(m-1)$.
This behavior is confirmed in Fig.~\ref{fig-observ}(b) where the
normalized entropy $S/N_K$ is plotted as a function of $\alpha$
for different small values of $D$: $S/N_K$ converges to a well-defined value
for $\alpha<-1$, while no clear convergence is observed for $\alpha>-1$.
Such a result can be understood in the framework of the simplified model
where $\Delta_M=\nu_0 K^{\alpha}$ and $\Delta_j=0$ for $j<M$.
For large $K$, $S$ is proportional to $N_K$ if $\alpha<-1$ and to
$N_K^{|\alpha|}$ if $-1<\alpha<0$, while $S$ is independent
of $N_K$ for $\alpha>0$.
Hence the dissipative state characterized by
$P_\mathrm{neq}(\{n_{j,i}\})$ has a much lower entropy than the
quasi-equilibrium state obtained for $\alpha<-1$.
In other words, the accessible volume in phase space is much smaller
in far-from-equilibrium states than in equilibrium states.

A major challenge for future work would be to characterize
such asymptotic nonequilibrium states in more realistic models,
and to understand their statistical fundations.
The fluctuation properties, and specifically the validity of
Gallavotti-Cohen relations \cite{Gallavotti95,Aumaitre01} in these states,
would be issues of great interest
\footnote{In the present model, the distribution of injected power
satisfies the Gallavotti-Cohen symmetry.}.

\end{document}